\documentclass{ptapap}

\author{Justyna {\'S}redzi{\'n}ska}[CAMK]
\author{Bo{\.z}ena Czerny}[CFT] 
\author{M.~Bilicki}[LO]
\author{K.~Hryniewicz}[CAMK]
\author{M.~Krupa}[UJ]
\author{A.~Kurcz}[UJ] 
\author{P.~Marziani}[PM]
\author{A.~Pollo}[NCBJ, UJ]
\author{W.~Pych}[CAMK] 
\author{A.~Udalski}[UW]
\affil[CAMK]{Nicolaus Copernicus Astronomical Center PAS, Bartycka 18, 00-716 Warsaw, Poland}
\affil[CFT]{Center for Theoretical Physics PAS, al.~Lotnik\'{o}w 32/46, 02-668 Warsaw, Poland}
\affil[LO]{Leiden Observatory, Leiden University, 2333CA Leiden, the Netherlands}
\affil[UJ]{Astronomical Observatory, Jagiellonian University, Orla 171, 30-244 Cracow, Poland}
\affil[NCBJ]{National Centre for Nuclear Research, Ho\.za 69, 00-681 Warsaw, Poland}
\affil[PM]{INAF, Osservatorio Astronomico di Padova, 35122 Padova, Italy}
\affil[UW]{Warsaw University Observatory, Al.~Ujazdowskie 4, 00-478 Warsaw, Poland}

\title{Tracing dark energy with quasars}

\begin{document}

\maketitle

\begin{abstract}

The nature of dark energy, driving the accelerated expansion of the Universe, is one of the most important issues in modern astrophysics. In order to understand this phenomenon, we need precise astrophysical probes of the universal expansion spanning wide redshift ranges. Quasars have recently emerged as such a probe, thanks to their high intrinsic luminosities and, most importantly, our ability to measure their luminosity distances independently of redshifts. Here we report our ongoing work on observational reverberation mapping using the time delay of the Mg II line, performed with the South African Large Telescope (SALT).

%The concept of dark energy was introduced in the process of understanding the evolution of the Universe. This is one of the most interesting topic in modern astronomy followed by the discovery of the accelerated expansion of the Universe. Precise measurement of this effect is a key to understand the nature of this medium, and we need good probes to do that. Quasars appears as an ideal candidate for this purpose as these objects are highly luminous and detected in wide range of redshift.

%From Big Bang to present time a lot of things happened and we are able to see amazing structures of galaxies and stars. In the beginning of Universe
%everything was blurred in space and the concept of dark energy was introduced in the process of understanding its evolution. The discovery of the accelerated expansion of the Universe
%gives us possibility to define new interesting topics in modern astronomy. Although there are some
%theoretical explanation for the existence of dark energy, yet it has remained the biggest puzzle among the astronomers and physicist.

\end{abstract}

\section{Introduction}

Dark energy (DE), estimated to constitute about 70\% of the universal mass-energy content today, is one of the crucial ingredients of the standard cosmological model. Since the first observational evidence of the accelerated expansion thanks to Supernovae Ia (SNeIa), various efforts have been undertaken to understand the nature of this phenomenon. On the observational side this requires designing ever more precise measurements, and various probes have been proposed and used to study DE. Quasars, currently detected at redshifts up to $z\sim7$, are ideally suited to this task thanks to their very high intrinsic luminosity and large number density. However, using them as standard candles in a similar way to SNeIa, requires knowing their individual absolute luminosities \citep{Watson2011}. Determining the latter is possible through the measurement of the time delay between the variable nuclear continuum and emission lines \citep{Czerny2013}, or by analysing the shape of the lines to measure the BLR size. Here we focus on the first of these methods, which is possible by measuring the delay of the H$\beta$ line, performed first for nearby Active Galactic Nuclei (AGN) \citep{Clavel1991, Peterson1993, Reichert1994, Chelouche2012, Bentz2013}. The time delays in quasars are of the order of a few years, so the observations require sparse monitoring over an extended period of time. %The second method, on the other hand, requires measurements of very large numbers of quasars to reduce the inevitable statistical error.

%Dark energy seems to be very important, when we following the cosmological models which estimate that it forms 70\% of the Universe content.
%Projects based on Supernovae type Ia brings us first observational proves for accelerating expansion of the Universe. Precise measurement of this effect is a 
%key to understand the nature of dark energy, and we need good probes to do that. With the development of observational techniques we can see deeper with better accuracy. 
%It was a matter of time to find sources which will allow for the spread of research regarding the existence of this medium. Quasars appears as an ideal candidate for 
%this purpose as these objects are numerous, highly luminous and detected in wide range of redshift (0<z<7). They can be used in similar way as standard candles only when the absolute 
%luminosity is determined for each of them apart, after \citep{Watson2011, Czerny2013}. This can be done through the measurement of the time delay between the variable nuclear continuum and the emission lines, or by 
%analysis of the shape of the emission line. The first method is confirmed by the delay measurement of H$\beta$ line done for nearby Active Galactic Nuclei (AGN) \citep{Clavel1991, Peterson1993, Reichert1994, 
%Chelouche2012, Bentz2013}. The time delays in quasars are of the order of a few years, so the observations requires sparse monitoring over an extended period of time. The second method 
%is under discussion, and numerous quasars are needed to reduce the inevitable statistical error.

\section{Method}
\begin{figure}
\centering
\includegraphics[width=0.4\textwidth]{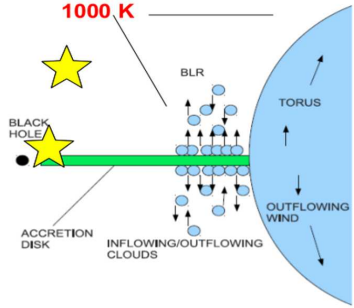}
\caption{\label{fig:blr}Broad line region as failed dusty wind; for details see \cite{Czerny2011}.}
\end{figure}

The work reported here is based on a simple mechanism of the broad line region (BLR) formation, presented in \cite{Czerny2011}, which explains how the size of BLR depends on the absolute monochromatic luminosity: 
%Work reported in this article is based on simple formation mechanism of the Broad Line Region (BLR) presented in \citep{Czerny2011} which explains size depends on the 
%absolute monochromatic luminosity. 
%We are able to derive this correlation from Shakura-Sunyaev accretion disk theory \citep{Shakura1973} and assumption that BLR formed in temprature 1000K.
$R_{\rm BLR} = {\rm const}\; L_v^{1/2}$.
In our method we assume that: (\textit{i}) we know the redshift of the source; (\textit{ii}) the optical/UV continuum is generated in the inner part of the accretion disk surrounding the central black hole, while the broad emission lines are produced in another disk region, in clouds above the accretion disk as shown in Fig.\ \ref{fig:blr}. Dust leads to outflow and forces the material to rise high above the disk, the dust cannot however survive in temperatures much higher than 1000~K. Strong radiation field destroys the dust (through evaporation) and the material falls back without a driving force. More accurately, this mechanism is proposed for the low ionization line part of the BLR, like H$\beta$ and MgII, which do not show a systematic shift in velocity with respect to the narrow line region (NLR). Using the observational method of reverberation mapping we measure the BLR size, which allows us to determine the absolute monochromatic luminosity of the quasar. Comparing the latter with the \textit{observed} monochromatic flux from photometry, we obtain the luminosity distance and are able to locate the source on the distance--redshift diagram.

%In our method we assume:  $(i)$ we know the redshift to the source, $(ii)$ optical/UV continuum comes from an inner part of accretion disk surrounding the central black hole and 
%Broad Emission Lines come from an other disk region, from clouds above an accretion disk shown in Figure \ref{fig:blr}. Dust leads to outflow and forces the material to rise high above 
%the disk but dust cannot survive in the temprature much higher than 1000K. Strong radiation field destroys the dust (evaporation process) and the material falls back without driving force.
%More accurately, this is proposed for the Low Ionization Line part of BLR, like H$\beta$ and MgII which do not show a systematic shift in velocity with respect to
%Narrow Line Region (NLR). Using observation-reverberation method we measure size of BLR what allows to determine the absolute monochromatic luminosity in quasar, its comparison to observed monochromatic flux 
%from photometry gives the luminosity distance and locates the source on distance - redshift diagram.

\begin{figure}
  \centering
%  \begin{minipage}{0.49\textwidth}
    \includegraphics[width=0.35\textwidth]{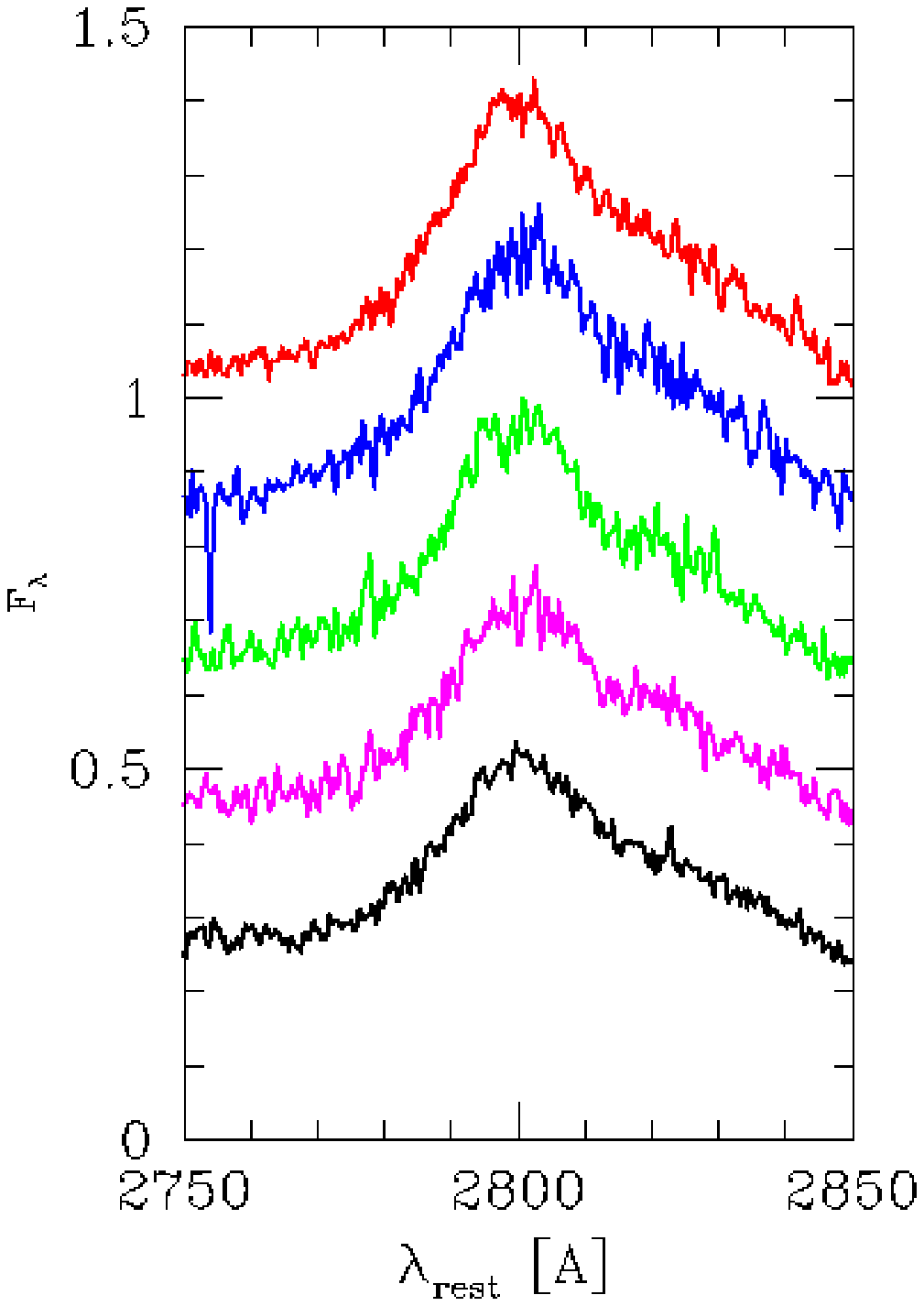}
%  \end{minipage}
%  \begin{minipage}{0.49 \textwidth}
\quad
    \includegraphics[width=0.35\textwidth]{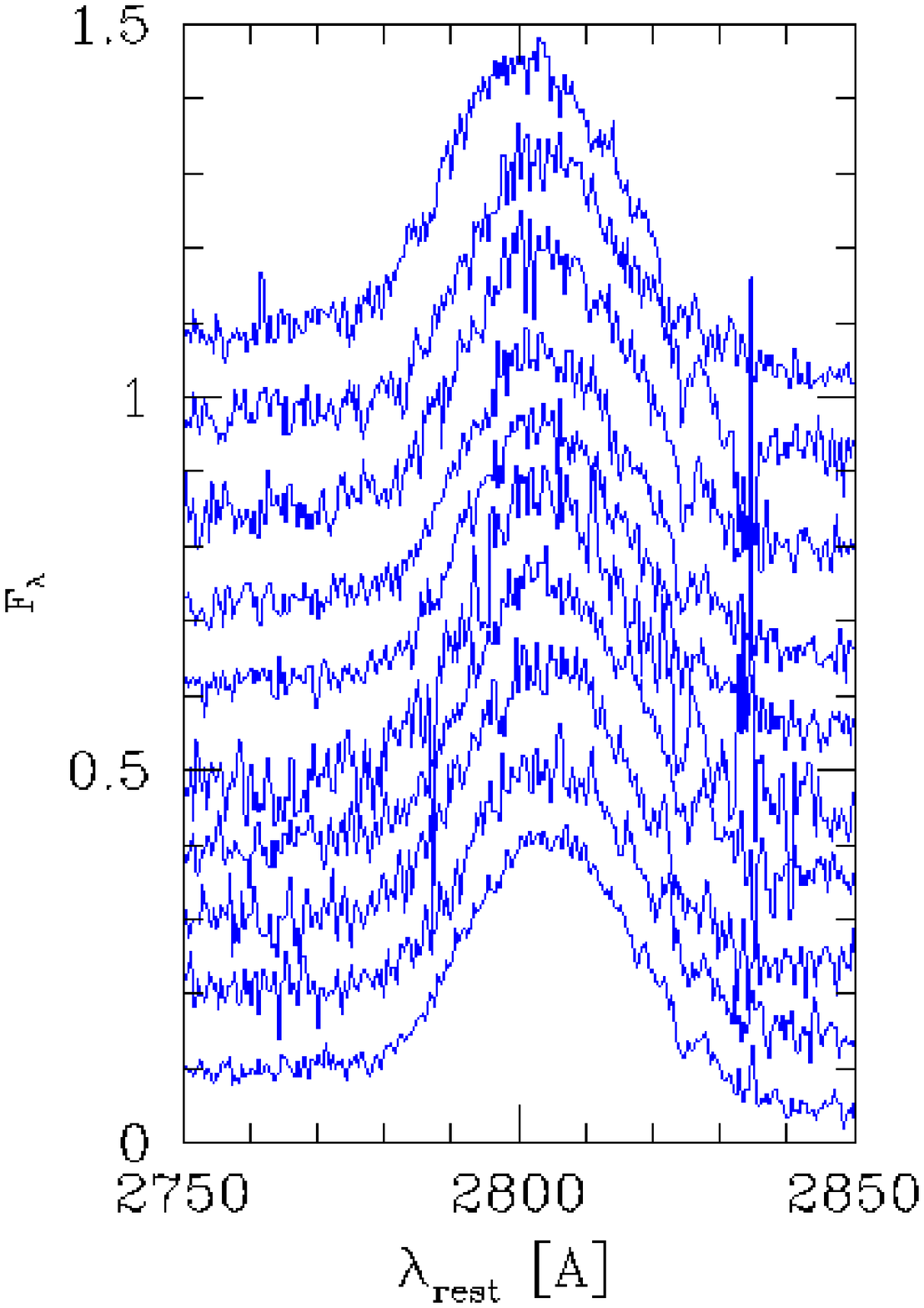}
%    \label{fig:he}
%  \end{minipage}
      \caption{\label{fig:cts}SALT spectroscopy of the MgII emission line for the quasar CTS~C30.10 (\textit{left}) and HE~0435-4312 (\textit{right}). The collected data cover respectively 15 months and 3 years of observations.}
\end{figure}

Until present, we have collected observations of three quasars at redshift $z\sim1$, and part of this work is already published \citep{Hryniewicz2014, Modzelewska2014}. Spectra of two sources, CTS~C30.10 and HE~0435-4312, are shown in Fig.\ \ref{fig:cts}. This project requires using the MgII line, and our studies are pioneer in this respect, as such monitoring has never been done before. Data obtained so far with the South African Large Telescope (SALT) show that we achieve the required accuracy (below 2\%) of the MgII measurement to determine its variability \citep{Modzelewska2014,Sredzinska2016}, and simulations indicate that the program can give the accuracy of 0.06--0.32 mag in the distance modulus for each of the concerned quasars \citep{Czerny2013}.

%We provide observations of tree quasars at redshift z$\sim$1 and part of this work is already published \citep{Hryniewicz2014, Modzelewska2014}, spectra for two sources CTS C30.10 and
%HE 0435-4312 are shown in Figure \ref{fig:cts} and \ref{fig:he}. This project requires using the MgII line, 
%and such monitoring has never been done before. Data performed so far with SALT showed that we achieve the requested accuracy - below 2 per cent - of the MgII measurement 
%to determine its variability, and simulations indicate that the program can give the accuracy of 0.06-0.32 magnitude in the distance modulus for each quasar concerned.

\section{Discussion}
Our method is very promising for DE studies, as it can easily reach to much higher redshifts than available with supernovae (ground-based observations of SNeIa hardly reach beyond $z = 1$). On the other hand, such reverberation studies require very large time spans, as at least 5 years of systematic observations are needed to estimate the emission line delay with respect to the continuum. High-quality data from SALT give us the opportunity to accurately model the emission line shape. This provides us with a new tool for cosmological analyses which aim at understanding the mystery of dark energy.

%This method is worth of studies, it can easily cover bigger redshift range than supernovae (the groudbased observations of SNe Ia do not extend much beyond z = 1). 
%On the other side, the reverberation studies are time consuming, we need at least 5 years of systematic observations to achive the goal and estimate emission line delay with 
%respect to the continuum. SALT data are high quality and gives oportunity to accurate modeling of the emission line shape. This provide us new tool for cosmological purpose and help to understand
%research regarding existence of dark energy.

\acknowledgements{This work was supported by Polish grants \#719/N-SALT/2010/0 and \#UMO-2012/07/B/ST9/04425. We acknowledge the support from the Foundation for Polish Science through the Master/Mistrz program 3/2012.}

\bibliographystyle{ptapap}
\bibliography{bibliografia}

\begin{thebibliography}{11}
\providecommand{\natexlab}[1]{#1}
\providecommand{\url}[1]{\texttt{#1}}
\providecommand{\urlprefix}{URL }
\providecommand{\eprint}[2][]{\url{#2}}

\bibitem[{{Bentz} et~al.(2013)}]{Bentz2013}
{Bentz}, M.~C., et~al., \emph{{The Low-luminosity End of the Radius-Luminosity
  Relationship for Active Galactic Nuclei}}, \emph{\apj} \textbf{767}, 149
  (2013), \eprint{1303.1742}

\bibitem[{{Chelouche} \& {Daniel}(2012)}]{Chelouche2012}
{Chelouche}, D., {Daniel}, E., \emph{{Photometric Reverberation Mapping of the
  Broad Emission Line Region in Quasars}}, \emph{\apj} \textbf{747}, 62 (2012),
  \eprint{1105.5312}

\bibitem[{{Clavel} et~al.(1991)}]{Clavel1991}
{Clavel}, J., et~al., \emph{{Steps toward determination of the size and
  structure of the broad-line region in active galactic nuclei. I - an 8 month
  campaign of monitoring NGC 5548 with IUE}}, \emph{\apj} \textbf{366}, 64
  (1991)

\bibitem[{{Czerny} \& {Hryniewicz}(2011)}]{Czerny2011}
{Czerny}, B., {Hryniewicz}, K., \emph{{The origin of the broad line region in
  active galactic nuclei}}, \emph{\aap} \textbf{525}, L8 (2011),
  \eprint{1010.6201}

\bibitem[{{Czerny} et~al.(2013)}]{Czerny2013}
{Czerny}, B., et~al., \emph{{Towards equation of state of dark energy from
  quasar monitoring: Reverberation strategy}}, \emph{\aap} \textbf{556}, A97
  (2013), \eprint{1212.0472}

\bibitem[{{Hryniewicz} et~al.(2014)}]{Hryniewicz2014}
{Hryniewicz}, K., et~al., \emph{{SALT long-slit spectroscopy of LBQS 2113-4538:
  variability of the Mg II and Fe II component}}, \emph{\aap} \textbf{562}, A34
  (2014), \eprint{1308.3980}

\bibitem[{{Modzelewska} et~al.(2014)}]{Modzelewska2014}
{Modzelewska}, J., et~al., \emph{{SALT long-slit spectroscopy of CTS C30.10:
  two-component Mg II line}}, \emph{\aap} \textbf{570}, A53 (2014),
  \eprint{1408.1520}

\bibitem[{{Peterson}(1993)}]{Peterson1993}
{Peterson}, B.~M., \emph{{Reverberation mapping of active galactic nuclei}},
  \emph{\pasp} \textbf{105}, 247 (1993)

\bibitem[{{Reichert} et~al.(1994)}]{Reichert1994}
{Reichert}, G.~A., et~al., \emph{{Steps toward determination of the size and
  structure of the broad-line region in active galactic nuclei. 5: Variability
  of the ultraviolet continuum and emission lines of NGC 3783}}, \emph{\apj}
  \textbf{425}, 582 (1994)

\bibitem[{{Sredzinska} et~al.(2016)}]{Sredzinska2016}
{Sredzinska}, J., et~al., \emph{{SALT long-slit spectroscopy of HE 0435-4312:
  fast change in the Mg II emission line shape}}, \emph{ArXiv e-prints}
  (2016), \eprint{1602.01975}

\bibitem[{{Watson} et~al.(2011){Watson}, {Denney}, {Vestergaard}, \&
  {Davis}}]{Watson2011}
{Watson}, D., {Denney}, K.~D., {Vestergaard}, M., {Davis}, T.~M., \emph{{A New
  Cosmological Distance Measure Using Active Galactic Nuclei}}, \emph{\apjl}
  \textbf{740}, L49 (2011), \eprint{1109.4632}

\end{thebibliography}

\end{document}